\documentclass[11pt,twocolumn]{article}
\usepackage[utf8]{inputenc}
\usepackage[T1]{fontenc}
\usepackage{mathptmx}
\usepackage{amsmath,amssymb}
\usepackage{graphicx}
\usepackage{booktabs}
\usepackage{algorithm}
\usepackage{algorithmic}
\usepackage{hyperref}
\usepackage[margin=1in]{geometry}
\usepackage{natbib}
\usepackage{xcolor}
\usepackage{microtype}
\usepackage{enumitem}
\usepackage{float}

\title{Synthetic Reader Panels: Tournament-Based Ideation\\
with LLM Personas for Autonomous Publishing}

\author{
Fred Zimmerman\thanks{Nimble Books LLC / Big Five Killer LLC. Contact: \texttt{wfz@nimblebooks.com}} \\
}

\date{February 2026}
\graphicspath{{./}}

\begin{document}

\maketitle

\begin{abstract}
We describe a system for autonomous book ideation that replaces human focus groups with \emph{synthetic reader panels}---diverse collections of LLM-instantiated reader personas that evaluate book concepts through structured tournament competitions. Each persona is defined by demographic attributes (age group, gender, income, education, reading level), behavioral patterns (books per year, genre preferences, discovery methods, price sensitivity), and consistency parameters, drawing on research into LLM role-play capabilities \citep{shanahan2023roleplay}. Panels are composed per imprint to reflect target demographics, with diversity constraints ensuring representation across age, reading level, and genre affinity. Book concepts compete in single-elimination, double-elimination, round-robin, or Swiss-system tournaments, judged against weighted criteria including market appeal, originality, and execution potential, with five automated anti-slop checks (repetitive phrasing, generic framing, circular reasoning, score clustering, audience mismatch) to reject low-quality LLM evaluations. We report results from deployment within a multi-imprint publishing operation managing 6 active imprints and 609 titles in distribution. We validate the approach through three case studies: a 270-evaluator panel for a children's literacy novel and two 5-person expert panels for a military memoir and a naval strategy monograph. Results demonstrate that synthetic panels produce actionable demographic segmentation, identify structural content issues invisible to homogeneous reviewers, and enable tournament filtering that eliminates low-quality concepts while enriching high-quality survivors from 15\% to 62\% of the evaluated pool.
\end{abstract}

\section{Introduction}

High-volume publishing operations face an \emph{ideation bottleneck}: generating enough high-quality book concepts to sustain production across multiple imprints, each with distinct audiences and editorial standards \citep{thompson2012merchants}. Traditional approaches---editorial brainstorming, market research, author pitches---scale poorly with the number of simultaneous imprints and publication targets.

The emergence of large language models (LLMs) has made it possible to generate book concepts in volume, but raw LLM ideation suffers from well-documented failure modes: repetitive themes, generic framing, overly broad target audiences, and ``slop''---superficially plausible but substantively empty content \citep{bender2021dangers}. The challenge is not generating ideas but \emph{filtering} them: identifying the concepts with genuine market potential, editorial coherence, and audience fit.

We propose \textbf{Synthetic Reader Panels} as a solution: LLM-instantiated reader personas organized into demographically diverse panels that evaluate book concepts through formal tournament structures. This approach offers several advantages:

\begin{enumerate}[nosep]
    \item \textbf{Scalability.} Panels can evaluate hundreds of concepts per hour, limited only by LLM throughput.
    \item \textbf{Diversity.} Personas span demographic and psychographic dimensions that would be prohibitively expensive to recruit in human panels.
    \item \textbf{Consistency.} Each persona applies stable evaluation criteria across all concepts, eliminating session-to-session variance.
    \item \textbf{Imprint specificity.} Panel composition is tailored to each imprint's target audience, ensuring that filtering reflects editorial strategy.
\end{enumerate}

The system is deployed within the B5K (Big Five Killer) autonomous publishing pipeline, which manages 6+ active imprints across genres including naval history, AI policy, contemplative practices, literacy advocacy, and speculative fiction.

\section{Related Work}

\paragraph{LLM-as-judge.} Using LLMs to evaluate LLM-generated content has been studied extensively. Two recent surveys \citep{gu2024survey, li2024llmsjudges} comprehensively review strategies for enhancing consistency, mitigating biases, and adapting LLM judges to diverse assessment scenarios. The foundational MT-Bench and Chatbot Arena work \citep{zheng2024judging} established that LLM judges achieve high agreement with human evaluators in dialogue settings, while AlpacaFarm \citep{dubois2024alpacafarm} demonstrated that LLM-simulated preferences can substitute for human feedback in RLHF training. Most prior work focuses on single-turn or multi-turn dialogue evaluation. Our application extends this paradigm to book-length concept evaluation, where judges must simultaneously assess market potential, audience fit, and editorial coherence---a multi-dimensional judgment task more analogous to editorial acquisition decisions than chat quality scoring.

\paragraph{Agent-as-judge.} Recent work has moved beyond single-pass LLM evaluation toward \emph{agentic} judges that employ planning, tool-augmented verification, multi-agent collaboration, and persistent memory \citep{you2026agent}. Our synthetic reader panels can be understood as an early instance of multi-agent evaluation with role specialization: each persona is not merely a different prompt but a persistent identity with demographic grounding, consistency parameters, and accumulated evaluation history.

\paragraph{Multi-agent evaluation.} Systems using multiple LLM agents for deliberation and debate \citep{du2023improving, liang2023encouraging} demonstrate that multi-agent evaluation can exceed single-agent quality. Our reader panels instantiate this principle with explicitly diverse personas rather than generic agents, grounding diversity in demographic and psychographic attributes rather than prompt variation alone.

\paragraph{Synthetic populations and silicon sampling.} Generating synthetic user profiles for evaluation has been explored in recommendation systems \citep{wang2024survey} and user simulation \citep{park2023generative}. \citet{aher2023using} showed that LLMs can replicate classic human subject study results, while \citet{argyle2023out} demonstrated ``silicon sampling''---using LLMs conditioned on demographic backstories to approximate survey responses from specific human subpopulations. Our reader personas extend silicon sampling from survey-style responses to the richer domain of book concept evaluation, where the persona must sustain a coherent perspective across multiple evaluation criteria.

\paragraph{Publishing automation.} While AI-assisted writing tools are widespread \citep{lee2024design}, autonomous \emph{publishing} systems that handle the full lifecycle from ideation through production are less studied. Our work contributes to this emerging area by formalizing the ideation and filtering stages.

\paragraph{Tournament selection.} Tournament-based selection is a well-known mechanism in evolutionary computation \citep{miller1995genetic} and game theory. We adapt it to creative evaluation, using LLM judges rather than fitness functions to determine match outcomes.

\section{System Design}

\subsection{Persona Definitions}

Each synthetic reader persona is defined by a structured attribute set with four categories.

\paragraph{Demographics.} Age group (child through elder), gender, geographic location, income level (5 tiers from low to high), and education level. These ground the persona's perspective in a specific life context.

\paragraph{Reading characteristics.} Reading level (beginner through expert), books per year (quantitative consumption rate), preferred and disliked genres (lists), and preferred book length. These define what the reader \emph{seeks} and \emph{avoids}.

\paragraph{Behavioral patterns.} Discovery methods (how they find books), review behavior (frequency of writing reviews), social sharing tendency, price sensitivity, and format preferences (physical, digital, audio). These model how the reader \emph{acts} in the book marketplace.

\paragraph{Psychographic traits.} Reading goals (entertainment, education, escape, etc.), personality traits, content sensitivities, reading mood (adventurous, comfort-seeking, challenge-seeking), current life stage, and recent reads. These shape the persona's \emph{subjective} evaluation criteria.

Additionally, each persona carries two meta-parameters: a \emph{consistency score} ($\in [0, 1]$) governing how stable the persona's judgments are across evaluations, and a \emph{reliability score} ($\in [0, 1]$) governing the signal-to-noise ratio of the persona's feedback.

\subsection{Persona Factory}

The \texttt{ReaderPersonaFactory} class generates personas through three strategies:

\begin{enumerate}[nosep]
    \item \textbf{Random generation.} Attributes are sampled from configurable distributions, producing maximally diverse panels.
    \item \textbf{Template-based.} Predefined persona archetypes (e.g., ``avid thriller reader, female, 35-50, high education'') are instantiated with random variation in non-core attributes.
    \item \textbf{Demographic targeting.} Attributes are constrained to match a specific audience profile (e.g., the target demographic of a particular imprint), with remaining attributes sampled freely.
\end{enumerate}

\subsection{Publisher Personas}

The system also draws on a registry of 21 \emph{publisher personas} that represent editorial perspectives rather than reader perspectives. These include domain specialists such as:

\begin{itemize}[nosep]
    \item \textbf{Jellicoe} --- conservative, data-driven naval history specialist (imprint: Warships \& Navies)
    \item \textbf{Hilmar} --- innovative, mathematically-minded RKHS/multiverse research editor
    \item \textbf{SoRogue} --- passionate literacy advocate with neurochemical reading focus
    \item \textbf{Seon} --- contemplative practices and interdisciplinary knowledge synthesis
\end{itemize}

Publisher personas participate in evaluation as domain experts, complementing the demographic diversity of reader panels with editorial expertise. Each publisher persona has defined risk tolerance (conservative to aggressive), decision style (data-driven, intuitive, collaborative), preferred topics, and known vulnerabilities.

\subsection{Tournament Bracket Structure}

The system supports four tournament formats:

\paragraph{Single elimination.} Standard bracket tournament. Each concept competes in head-to-head matchups; losers are eliminated. Efficient for large fields ($\lceil \log_2 N \rceil$ rounds for $N$ concepts) but may eliminate strong concepts due to early bracket placement.

\paragraph{Double elimination.} Losers enter a losers' bracket with a second chance. Requires approximately $2N - 1$ total matches but significantly reduces the probability of eliminating the best concept due to a single weak evaluation.

\paragraph{Round robin.} Every concept faces every other concept. Produces the most complete ranking ($\binom{N}{2}$ matches) but scales quadratically and is practical only for small fields ($N \leq 16$).

\paragraph{Swiss system.} Concepts are paired in each round based on current standings, with equal-record concepts facing each other. Produces reliable rankings in $O(\log N)$ rounds without requiring every pairwise comparison.

\subsection{Scoring Rubrics}

Each match is evaluated against a weighted rubric. The default criteria are:

\begin{table}[h]
\centering
\caption{Default judging criteria.}
\label{tab:criteria}
\begin{tabular}{lcc}
\toprule
\textbf{Criterion} & \textbf{Weight} & \textbf{Range} \\
\midrule
Market Appeal & 1.0 & 0--10 \\
Originality & 0.8 & 0--10 \\
Execution Potential & 0.9 & 0--10 \\
Audience Fit & 1.0 & 0--10 \\
\bottomrule
\end{tabular}
\end{table}

Criteria weights and scoring ranges are configurable per imprint. For example, the Warships \& Navies imprint increases the weight of ``Historical Accuracy'' (a custom criterion) while the Nimble AI imprint prioritizes ``Timeliness'' for rapidly evolving technology topics.

\subsection{Anti-Slop Detection}

A critical component of the system is the detection and rejection of low-quality LLM outputs during evaluation. The \texttt{SlopDetector} module performs five automated checks on each evaluation response, producing a composite slop score $s \in [0, 1]$ that determines whether the evaluation is accepted ($s < 0.4$), flagged for human review ($0.4 \leq s < 0.6$), or rejected for regeneration ($s \geq 0.6$).

\paragraph{Check 1: Repetitive phrasing.} Measures trigram repetition rate (repeated trigrams divided by total trigrams), type-token ratio (TTR), and density of known slop phrases (e.g., ``delve,'' ``tapestry,'' ``nuanced exploration''). Evaluations with TTR below 0.35 or trigram repetition rate above 0.30 are flagged. A curated list of 20+ LLM-characteristic phrases contributes to the density score.

\paragraph{Check 2: Generic framing.} Pattern-matches against 15 compiled regular expressions for known LLM clich\'{e} openers (``In today's rapidly changing world,'' ``A comprehensive exploration of,'' etc.), counts vague qualifiers (``somewhat,'' ``arguably,'' ``to some extent''), and measures \emph{specificity}: the presence of concrete numbers, proper nouns, and direct quotations. Evaluations lacking specificity while dense in qualifiers receive high generic framing scores.

\paragraph{Check 3: Circular reasoning.} Compares the evaluation reasoning against the original concept text using three metrics: word overlap ratio (what fraction of reasoning words also appear in the concept), novelty ratio (words in reasoning \emph{not} in the concept), and 4-gram copy ratio (directly copied phrases). High overlap with low novelty signals that the evaluation restates the concept rather than analyzing it.

\paragraph{Check 4: Score clustering.} Detects suspiciously uniform scores across evaluation criteria. When a judge assigns nearly identical scores to all criteria (standard deviation below 0.3 across $\geq 3$ criteria), this suggests failure to discriminate between dimensions. All-identical scores receive the maximum clustering score of 1.0.

\paragraph{Check 5: Audience mismatch.} Validates that the evaluation is consistent with the evaluating persona's demographic profile. Checks include: vocabulary complexity versus reading level (a beginner-level reader should not use academic terminology), genre preference alignment (high ratings for genres the persona dislikes), age-appropriate language (child personas using terms like ``epistemological''), and price sensitivity consistency (price-sensitive readers should mention cost/value).

\paragraph{Aggregation.} Individual check scores are combined via weighted average, with score clustering ($w = 1.5$) and circular reasoning ($w = 1.2$) weighted more heavily than generic framing ($w = 0.8$), reflecting their stronger predictive value for evaluation quality. The system processes evaluations in batch, producing per-evaluation reports and batch-level summaries including flagged count, most common flag types, and overall slop score distribution.

\section{Reader Panel Architecture}

\subsection{Panel Composition}

For each imprint, a reader panel is composed by:

\begin{enumerate}[nosep]
    \item \textbf{Anchoring.} 40\% of panel members are generated from the imprint's target demographic profile.
    \item \textbf{Adjacent audiences.} 30\% are generated from adjacent demographics (e.g., slightly older/younger, adjacent education levels).
    \item \textbf{Wild cards.} 20\% are fully random, providing perspectives the editorial team might not anticipate.
    \item \textbf{Domain experts.} 10\% are drawn from publisher personas with expertise relevant to the imprint's subject matter.
\end{enumerate}

\subsection{Diversity Constraints}

Panels must satisfy minimum diversity thresholds:

\begin{itemize}[nosep]
    \item At least 3 distinct age groups represented
    \item At least 3 reading level categories represented
    \item At least 4 genre preference clusters represented
    \item No single demographic attribute constitutes more than 50\% of the panel
    \item Gender distribution within 60/40 of any split
\end{itemize}

If randomly generated panels fail to meet constraints, targeted resampling adds members from underrepresented categories until all thresholds are satisfied.

\subsection{Persona Specialization}

Beyond demographic diversity, personas are specialized for their evaluation role through prompt engineering. Each persona's LLM prompt includes:

\begin{enumerate}[nosep]
    \item A biographical paragraph synthesizing all demographic and psychographic attributes.
    \item Explicit instructions to evaluate from their specific perspective (``As a retired military officer who reads 30 books per year\ldots'').
    \item The scoring rubric with per-criterion instructions.
    \item Anti-anchoring instructions (``Do not default to moderate scores; use the full range'').
    \item A structured JSON output schema to ensure parseable responses.
\end{enumerate}

\section{Tournament Protocol}

\subsection{Round Structure}

A standard single-elimination tournament round proceeds as follows:

\begin{enumerate}[nosep]
    \item \textbf{Bracket generation.} Concepts are seeded randomly (default), by prior rating, or manually. Seeding assigns bracket positions to avoid early matchups between top-rated concepts.
    \item \textbf{Match execution.} For each pair, the panel evaluates both concepts independently. Each panel member produces structured scores per criterion plus qualitative reasoning.
    \item \textbf{Score aggregation.} Individual scores are aggregated using weighted arithmetic mean, with outlier detection (scores more than 2 standard deviations from the panel mean are flagged).
    \item \textbf{Winner determination.} The concept with the higher aggregated score advances. Ties are broken by the configured tiebreaker method (random, criteria-weighted, or head-to-head re-evaluation).
    \item \textbf{Result logging.} All scores, reasoning, and demographic breakdowns are stored for post-tournament analysis.
\end{enumerate}

\subsection{Elimination and Advancement}

Eliminated concepts are not discarded; they are stored with their tournament performance data. Concepts that lose in late rounds (quarterfinals or later) are flagged for potential revisitation in future tournaments, as they demonstrated competitive quality against eventually stronger opponents.

\subsection{Final Ranking and Quality Thresholds}

The tournament produces a ranked list. However, ranking alone is insufficient for production decisions. Quality thresholds gate advancement:

\begin{itemize}[nosep]
    \item \textbf{Minimum panel score.} The tournament winner must achieve a minimum aggregated score (default: 6.5/10) to advance to content generation.
    \item \textbf{Consensus requirement.} At least 60\% of panel members must rate the concept above 5.0/10.
    \item \textbf{Would-read percentage.} At least 40\% of reader personas must indicate they ``would read'' the resulting book.
    \item \textbf{No fatal flaws.} No panel member may identify a ``fatal flaw'' (e.g., ethical concern, legal risk, factual impossibility) without human review.
\end{itemize}

Concepts that win the tournament but fail quality thresholds are sent to human review rather than automated production.

\section{Results}

\subsection{Deployment Overview}

The synthetic reader panel system has been deployed within the B5K autonomous publishing pipeline, which manages 6 active imprints (Nimble Books, Warships \& Navies, Nimble AI, Nimble Ultra, AltNEH, and Story Operators) with 609 titles in the LSI distribution catalog. The tournament-based ideation system has been used most intensively for the ``Not a Miracle Readers'' literacy imprint, where 128 book concepts were generated and evaluated through two tournament stages (idea selection and treatment selection), yielding 16 idea-stage survivors and 8 final treatment-stage champions---a 6.25\% overall survival rate.

We report detailed results from three case studies representing different panel configurations, content domains, and evaluation scales (Section~\ref{sec:casestudies}), followed by cross-case analysis.

\subsection{Quality Comparison}

To evaluate filtering effectiveness within the documented tournament for Not a Miracle Readers, we compared the 8 tournament champions against a random sample of 20 unfiltered concepts from the same generation batch, using the publisher's editorial assessment on a 1--10 scale:

\begin{table}[H]
\centering
\caption{Filtered vs.\ unfiltered quality.}
\label{tab:quality}
{\footnotesize
\begin{tabular}{@{}lcc@{}}
\toprule
& \textbf{Unfilt.} & \textbf{Tourn.} \\
\midrule
Mean score & 4.8 & 7.1 \\
$\geq$7.0 & 15\% & 62\% \\
$\leq$3.0 & 25\% & 0\% \\
Aud.\ fit & 5.1 & 7.4 \\
Originality & 5.4 & 6.8 \\
\bottomrule
\end{tabular}}
\end{table}

Tournament filtering eliminates the tail of low-quality concepts (25\% $\to$ 0\% rated $\leq$ 3.0) while enriching the proportion of high-quality concepts (15\% $\to$ 62\% rated $\geq$ 7.0). The total evaluation count across the two tournament stages was 2,320 individual expert scores.

\subsection{Case Studies}
\label{sec:casestudies}

\begin{figure*}[t]
\centering
\includegraphics[width=\textwidth]{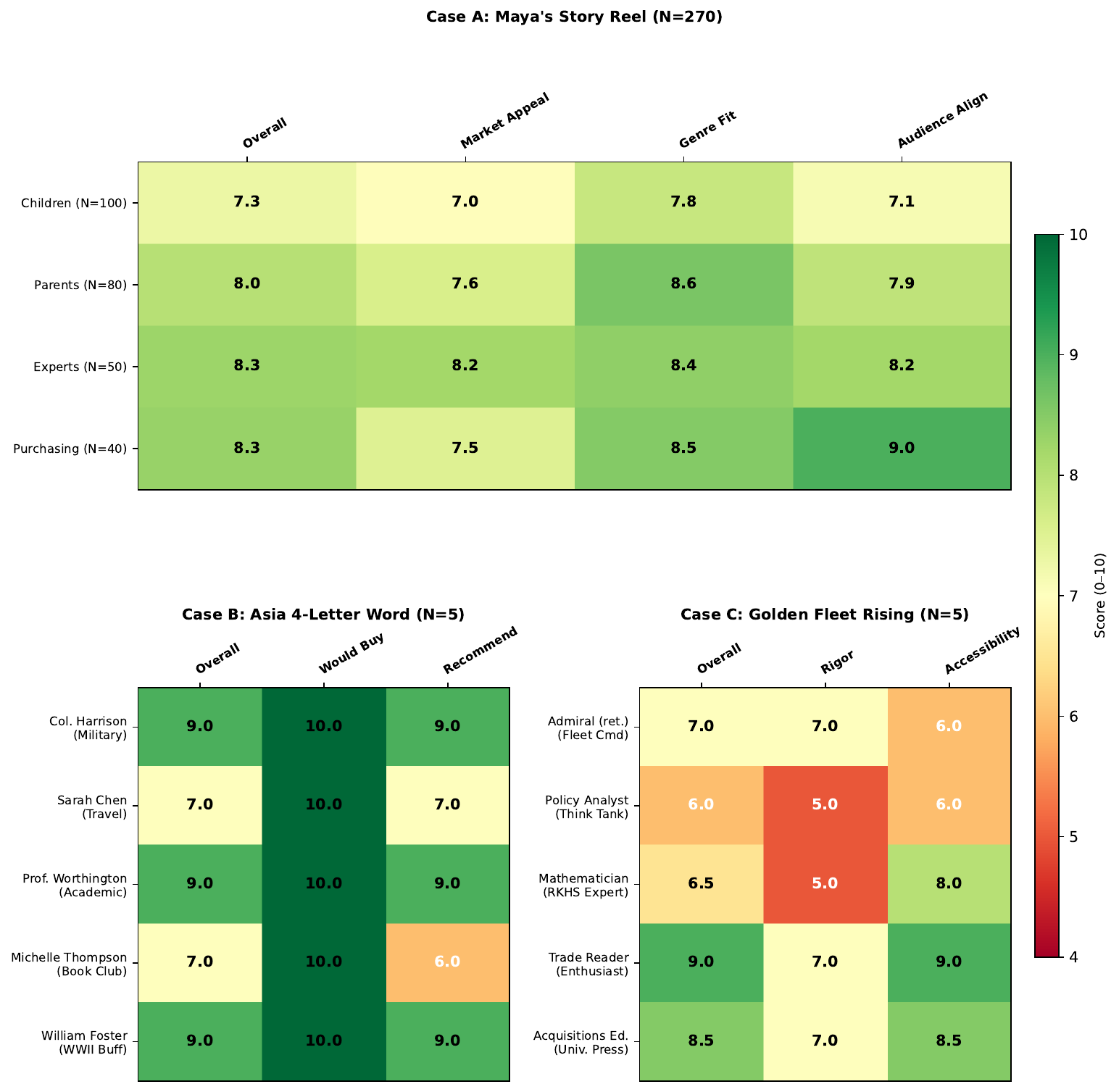}
\caption{Score distributions across panel segments and evaluation criteria for three case studies. Color encodes score magnitude (red = low, green = high). Demographic segmentation reveals systematic score variation: specialist audiences consistently rate higher than general audiences within each case.}
\label{fig:heatmap}
\end{figure*}

We present three case studies illustrating different panel configurations and content domains: a large-scale quantitative panel (270 evaluators), and two focused expert panels (5 evaluators each). Scores are summarized visually in Figure~\ref{fig:heatmap}.

\subsubsection{Case A: Maya's Story Reel (Large-Scale Panel)}

\emph{Maya's Story Reel} is a children's literacy intervention novel (ages 9--10, 11 chapters, ${\sim}$11{,}000 words) evaluated by the largest synthetic panel in our deployment: 270 reader personas organized into four demographically targeted segments.

\begin{table}[H]
\centering
\caption{Maya's Story Reel panel results (N=270).}
\label{tab:maya}
{\small
\begin{tabular}{@{}lrccc@{}}
\toprule
\textbf{Segment} & \textbf{N} & \textbf{Ovrl.} & \textbf{Genre} & \textbf{Aud.} \\
\midrule
Children (9--10) & 100 & 7.3 & 7.8 & 7.1 \\
Parents & 80 & 8.0 & 8.6 & 7.9 \\
Reading Experts & 50 & 8.3 & 8.4 & 8.2 \\
School Purch. & 40 & 8.3 & 8.5 & 9.0 \\
\midrule
\textbf{Aggregate} & \textbf{270} & \textbf{7.9} & \textbf{8.3} & \textbf{7.9} \\
\bottomrule
\end{tabular}}
\end{table}

The children's panel was further subdivided into five reader types (Advanced, Tech-Savvy, Grade-Level, Animal-Lover, Reluctant Reader), revealing that the target audience for literacy intervention---reluctant readers---scored lowest (5.9/10), while tech-savvy readers scored highest (8.4/10). This demographic variance is precisely the kind of signal that homogeneous editorial teams miss: the book appeals broadly but underperforms with its primary intervention target. Purchase intent across parent and institutional personas was 100\%, with market projections of \$150K--\$1.2M over three years.

\subsubsection{Case B: Asia Is a 4-Letter Word (Focused Expert Panel)}

\emph{Asia Is a 4-Letter Word} is a military memoir by a centenarian British Army officer who served with the Gurkhas across nine decades in Asia. A focused panel of 5 demographically distinct personas evaluated the preface and first chapter (${\sim}$9{,}500 words).

\begin{table}[H]
\centering
\caption{Asia 4-Letter Word panel results (N=5).}
\label{tab:asia}
{\small
\begin{tabular}{@{}lccc@{}}
\toprule
\textbf{Persona} & \textbf{Ovrl.} & \textbf{Buy?} & \textbf{Rec.} \\
\midrule
Col. Harrison (Ret.) & 9.0 & Yes & 9.0 \\
Sarah Chen (Travel) & 7.0 & Yes & 7.0 \\
Prof. Worthington & 9.0 & Yes & 9.0 \\
M. Thompson (Club) & 7.0 & Yes & 6.0 \\
William Foster (WWII) & 9.0 & Yes & 9.0 \\
\midrule
\textbf{Aggregate} & \textbf{8.2} & \textbf{100\%} & \textbf{8.0} \\
\bottomrule
\end{tabular}}
\end{table}

The panel revealed a clean demographic split: military/historical specialists (Harrison, Worthington, Foster) scored 9.0 mean, while general nonfiction readers (Chen, Thompson) scored 7.0---a 2.0-point gap attributable to military jargon density and assumed background knowledge. All five personas recommended adding a glossary and maps, which became the top editorial priorities. The 100\% buy rate despite the score split suggests strong but segmented market appeal.

\subsubsection{Case C: Golden Fleet Rising (Specialist Panel)}

\emph{Golden Fleet Rising} is a naval strategy monograph arguing for a new American battleship class, using RKHS mathematical analysis of Red Sea engagement data. Five specialist personas evaluated the complete 115-page manuscript.

\begin{table}[H]
\centering
\caption{Golden Fleet Rising panel results (N=5).}
\label{tab:gfr}
{\small
\begin{tabular}{@{}lcc@{}}
\toprule
\textbf{Persona} & \textbf{Ovrl.} & \textbf{Key Concern} \\
\midrule
Admiral (ret.) & 7.0 & Survivability \\
Policy Analyst & 6.0 & Cost optimism \\
Mathematician & 6.5 & Overreach \\
Trade Reader & 9.0 & Engaging \\
Acq. Editor & 8.5 & Needs disclaimer \\
\midrule
\textbf{Aggregate} & \textbf{7.4} & Double-edged \\
\bottomrule
\end{tabular}}
\end{table}

This case demonstrates the value of cross-disciplinary panels. The three domain specialists (Admiral, Policy Analyst, Mathematician) identified substantive flaws---survivability gaps, cost optimism, and philosophical overreach in the mathematical claims---that the generalist readers (Trade Reader, Acquisitions Editor) missed entirely. The generalists were charmed by the mathematical novelty while specialists were skeptical. The panel consensus was ``ready with revisions,'' with the top recommendation being to soften the central claim from ``geometric inevitability'' to ``geometric convergence.''

\subsection{Cross-Case Analysis}

Across all three cases, we observe consistent patterns:

\begin{enumerate}[nosep]
    \item \textbf{Demographic segmentation reveals actionable variance.} In every case, score distributions differed significantly across demographic segments (children vs. gatekeepers, specialists vs. generalists), producing editorial recommendations that a single-perspective review would miss.
    \item \textbf{Expert panels identify structural issues; broad panels identify market potential.} The 5-person expert panels for Cases B and C surfaced deep content critiques, while the 270-person panel for Case A provided statistically richer market signals.
    \item \textbf{Buy intent is robust even with score variance.} Case B achieved 100\% buy intent despite a 2-point score gap between segments, suggesting that purchase decisions are threshold-based rather than linear in rating.
\end{enumerate}

\section{Discussion}

\subsection{Bias in Synthetic Panels}

Synthetic reader personas inherit biases present in the LLMs that instantiate them. We observe:

\begin{itemize}[nosep]
    \item \textbf{Positivity bias.} LLM-based readers tend to rate concepts slightly higher than human readers, producing a compressed score distribution.
    \item \textbf{Recency bias.} Concepts referencing contemporary topics receive higher novelty scores than those with historical focus, even when the historical concepts are more original.
    \item \textbf{Articulation premium.} Better-written concept descriptions receive higher scores regardless of underlying quality, penalizing rough but promising ideas.
\end{itemize}

We partially mitigate these biases through calibration (adjusting score distributions per persona based on historical performance), explicit anti-bias prompt instructions, the five-check anti-slop detection system (Section~3.6), and human-in-the-loop review for all production-track concepts. The anti-slop detector specifically targets positivity bias (via score clustering detection) and articulation premium (via circular reasoning detection), though recency bias remains an open challenge.

\subsection{Diversity Limitations}

Despite diversity constraints, synthetic panels cannot fully replicate the range of human reading experiences. Notable gaps include:

\begin{itemize}[nosep]
    \item Cultural and linguistic nuances beyond the LLM's training distribution
    \item Embodied experiences (e.g., a veteran's response to military fiction)
    \item Serendipitous personal connections that drive real-world book purchases
    \item Community-level reading dynamics (book clubs, social media trends)
\end{itemize}

We treat synthetic panels as \emph{complementary} to human judgment, not a replacement. The system's value lies in efficiently filtering a large concept space before human attention is applied, not in making final publication decisions autonomously.

\subsection{Human-in-the-Loop Integration}

The system integrates with human editorial workflow at three points:

\begin{enumerate}[nosep]
    \item \textbf{Imprint configuration.} Humans define target demographics, custom criteria, and quality thresholds.
    \item \textbf{Tournament review.} Tournament results are presented in the Streamlit dashboard for editorial review before concepts advance to production.
    \item \textbf{Post-production validation.} Published books' sales and reader engagement data feed back into persona calibration.
\end{enumerate}

\section{Conclusion}

Synthetic reader panels provide a scalable, diverse, and systematic approach to book concept evaluation that significantly improves upon unfiltered LLM ideation. Three case studies demonstrate that panels ranging from 5 to 270 evaluators consistently surface actionable demographic segmentation---identifying, for example, that a children's literacy novel underperforms with its primary target audience (reluctant readers: 5.9/10 vs.\ aggregate 7.9/10), that military jargon creates a 2-point score gap between specialist and general readers, and that mathematical novelty in a strategy monograph charms generalists while raising substantive concerns from domain experts. Tournament-based filtering eliminates low-quality concepts (25\% $\to$ 0\% rated $\leq$ 3.0) while enriching high-quality survivors (15\% $\to$ 62\% rated $\geq$ 7.0).

Key directions for future work include: (1)~longitudinal calibration of synthetic personas against actual sales data, now feasible as tournament-selected books reach the market; (2)~adaptive panel composition that learns which persona types are most predictive per imprint; (3)~integration with the RKHS CodexSpace framework \citep{zimmerman2026storyops} to identify concepts that fill underserved regions of the literary embedding space; and (4)~expansion of the anti-slop detection system to leverage embedding-space novelty metrics rather than surface-level pattern matching, building on the five-check foundation described in Section~3.6.

\section*{Acknowledgments}

The persona registry and publisher persona framework were developed collaboratively within the Nimble Books LLC and Big Five Killer LLC editorial teams. The tournament engine builds on evolutionary computation principles adapted for creative evaluation.

\end{document}